\documentclass[twocolumn,aps,prl,superscriptaddress,amsmath,amssym,floatfix,nofootinbib]{revtex4-1}
\usepackage{lipsum}
\usepackage{graphicx}
\usepackage{bm}
\usepackage{dsfont} 
\usepackage{amsmath}
\usepackage{color}
\usepackage[usenames,dvipsnames]{xcolor}
\usepackage{tikz}
\usetikzlibrary{%
    decorations.pathreplacing,%
    decorations.pathmorphing,%
    arrows
}
\usepackage{amssymb}
\usepackage{enumerate}
\usepackage{comment}
\usepackage{listings}
 \usepackage{latexsym} 
\usepackage{revsymb} 
\usepackage{accents}
\usepackage{color}
\usepackage[colorlinks=true,linkcolor=MidnightBlue,citecolor=blue,filecolor=green,urlcolor=PineGreen]{hyperref}

\def\la{\langle}
\def\ra{\rangle}

\def\be{\begin{equation}}
\def\ee{\end{equation}}

\definecolor{codegreen}{rgb}{0,0.6,0}
\definecolor{codegray}{rgb}{0.5,0.5,0.5}
\definecolor{codepurple}{rgb}{0.58,0,0.82}
\definecolor{backcolour}{rgb}{0.95,0.95,0.95}
\definecolor{internationalorange}{rgb}{1.0, 0.31, 0.0}
\definecolor{cadetgrey}{rgb}{0.57, 0.64, 0.69}

\lstdefinestyle{mystyle}{
    backgroundcolor=\color{backcolour},   
    commentstyle=\tt\color{Cerulean},
    keywordstyle=\tt\color{Fuchsia},
    numberstyle=\it\tiny\color{MidnightBlue},
    stringstyle=\tt\color{red},
    basicstyle=\tt\footnotesize,
    breakatwhitespace=false,         
    breaklines=true,                 
    captionpos=b,                    
    keepspaces=true,                 
    numbers=left,                    
    numbersep=5pt,                  
    showspaces=false,                
    showstringspaces=false,
    showtabs=false,                  
    tabsize=2
}
 
\begin{document}

\title{Enhanced weak-value amplification via photon recycling}
\author{Courtney Krafczyk}
\affiliation{Department of Physics, University of Illinois at Urbana-Champaign, Urbana, IL 61801, USA}
\author{Andrew N. Jordan} 
\affiliation{Department of Physics and Astronomy, University of Rochester, Rochester, NY 14627, USA}
\affiliation{Institute for Quantum Studies, Chapman University, Orange, CA 92866, USA}

\author{Michael E. Goggin}
\affiliation{Department of Physics, Truman State University, Kirksville, MO 63501, USA}
\author{Paul G. Kwiat}
\affiliation{Department of Physics, University of Illinois at Urbana-Champaign, Urbana, IL 61801, USA}
\date{\today}

\begin{abstract} 
 In a quantum-noise limited system, weak-value amplification using post-selection normally does not produce more sensitive measurements than standard methods for ideal detectors: the increased weak value is compensated by the reduced power due to the small post-selection probability. Here we experimentally demonstrate {\it recycled} weak-value measurements using a pulsed light source and optical switch to enable nearly deterministic weak-value amplification of a mirror tilt. Using photon counting detectors, we demonstrate a signal improvement by a factor of $4.4 \pm 0.2$ and a signal-to-noise ratio improvement of $2.10 \pm 0.06$, compared to a single-pass weak-value experiment, and also compared to a conventional direct measurement of the tilt.  The signal-to-noise ratio improvement could reach around 6 for the parameters of this experiment, assuming lower loss elements.
\end{abstract}

\maketitle

{\it Introduction---}
Weak-value \cite{aharonov1988result} amplification has been successfully implemented in a variety of optical platforms to sensitively measure a number of system parameters \cite{dressel2014colloquium}.  This method has measured the optical spin Hall shift of 1 \AA\,  \cite{hosten2008observation}, 4-pm displacement or 400-frad angular tilt measurements \cite{dixon2009ultrasensitive}, 
frequency measurements of 130 kHz \cite{starling2010precision},
velocity measurements of 400 fm/s\, \cite{viza2013weak},
temperature shifts of 0.2 mK precision \cite{egan2012weak}, glucose concentration of $9 \times 10^{-5}$ g/L\, \cite{li2016application}, magnetic field sensitivities of 7 fT \cite{qu2018sub}, simultaneous multiparameter measurement \cite{vella2019simultaneous}, 
fine-tuned beam displacements \cite{salazar2015demonstration}, among many other experiments.  The method is inspired by a quantum effect where the shift of a quantum meter is amplified by the weak value of an operator, but with the sacrifice of the count rate by the probability of postselection on a given result of a subsequent measurement.  In an ideal, quantum-limited situation using coherent laser light, these two effects balance each other for the purposes of making precision measurements \cite{starling2009optimizing}.  However, in many practical situations there are other noise sources this technique {\it can} suppress, such as time-correlated noise \cite{feizpour2011amplifying,sinclair2017weak}, systematic noise \cite{pang2016protecting,li2018phase}, and other assorted sources such as jitter and turbulence \cite{viza2015experimentally,jordan2014technical,lyons2018noise}.  
These advantages of implementing weak value amplification are all in spite of the fact that the vast majority of events are discarded.  The postselected events can give equal performance in the quantum-limited case because the available information about the parameter of interest is concentrated into the few measured events, so the discarded events contain negligible information \cite{jordan2014technical}.  Nevertheless, in an optical context, the discarded events are photons that could still be used as a resource.  Indeed, in metrology, the resource that is used is typically quantified as the number of photons used in the experiment.  To this end, we can obtain a further metrological advantage by recycling these photons by reinjecting them back into the system so they are not wasted.

\begin{figure}[tbh]
    \includegraphics[width=8cm]{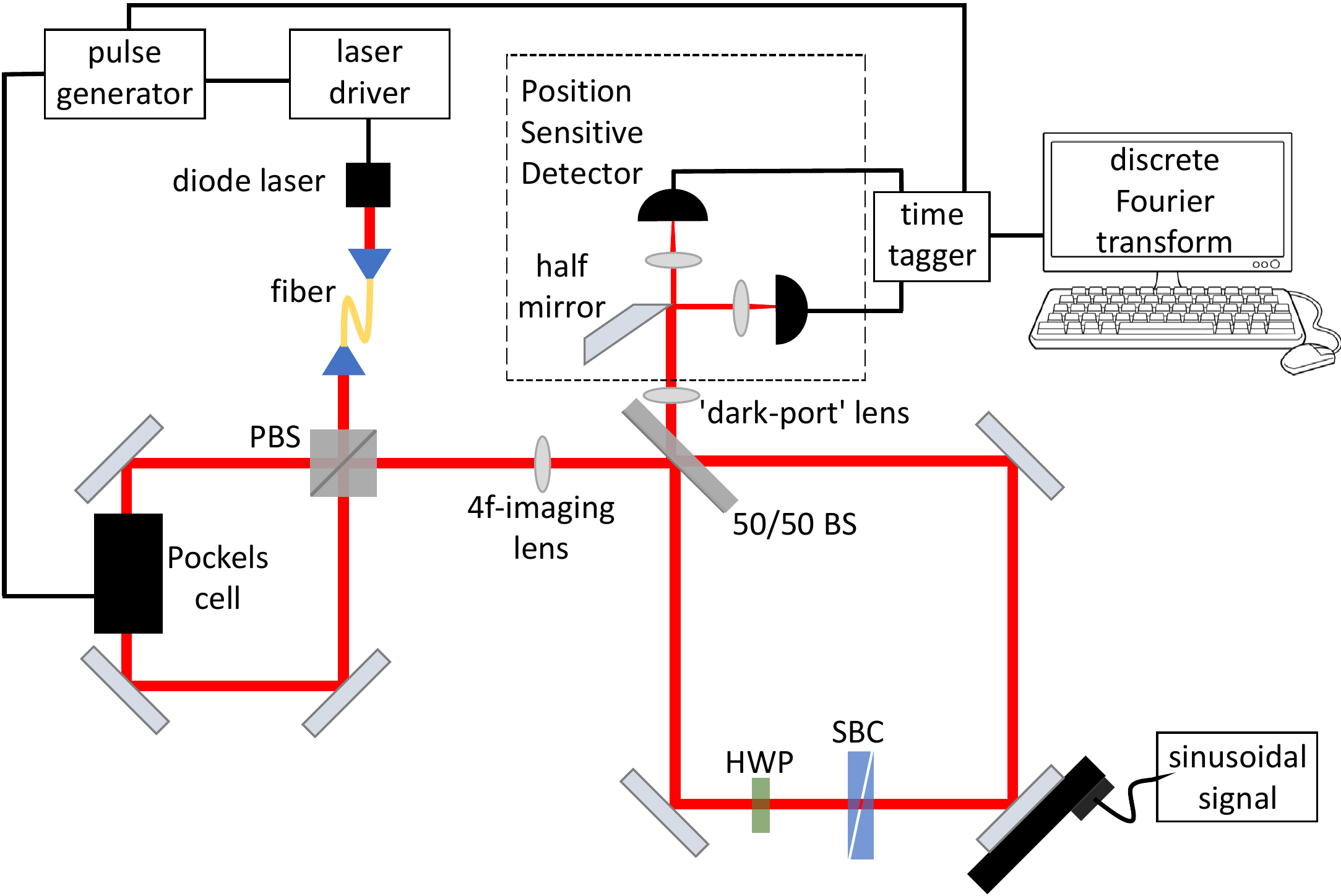}
    \caption{Schematic of the experiment.  The Sagnac interferometer in the lower right part of the apparatus implements weak value amplification.  It consists of the 50/50 beam splitter, a half wave plate (HWP) and a Soleil Babinet compensator (SBC) to introduce a relative phase shift between the clockwise and counter-clockwise light paths. Once the light pulse passes into the interferometer, the Pockels cell (lower left) fires, so the combination of the polarizing beam splitter and the loop on the lower left part of the apparatus re-injects any light that is not amplified and measured at the Position Sensitive Detector (PSD), implemented with a ``knife-edge" mirror and two avalanche photodiodes. Photon detection events are time tagged to investigate the arrival time statistics. Note: Position of PSD and `dark-port' lens not to scale.}
    \label{fig:setup}
\end{figure}

In this paper, we implement the proposal of Dressel {\it et al.} \cite{dressel2013strengthening}, using a Pockels cell and polarization optics to realize photon recycling with a pulsed light source. To explore the quantum limit of the technique, our experiment is carried out at the single-photon level, in contrast to most previous weak value-based metrology experiments (with few exceptions \cite{Pryde2005,Hallaji2017}). As shown in Fig. 1, the interferometer is designed to implement weak-value amplification on the light exiting the system, so that a small tilt of the interferometer mirror results in an amplified deflection of the light exiting the dark port (see Ref. \cite{barnett2003ultimate} for a discussion of deflection measurements).  The Pockels cell is designed to fire after the input photon passes, so that after the polarized pulse leaves the interferometer bright port, the combined effect of the polarizing beam splitter and the Pockels cell (which rotates the polarization by 90$^{\circ}$ when switched on) reinjects the pulse back into the interferometer.  This process continues, at least in principle, until {\it all} the light exits the dark port, each photon experiencing the weak-value amplification.  This results in the signal-to-noise of the measurement being weak-value amplified by the photon recycling.  A two-round recycle was demonstrated in Ref.~\cite{byard2014increasing}.


{\it Theory---} Weak-value amplification is by now a well understood and increasingly commonly applied metrological technique.  Consisting of a system and a meter, a continuous meter degree of freedom $x$ is displaced by the amount $g A_w$, where $g$ is the coupling constant between system and meter, and 
\be
A_w \equiv \la f |A|i\ra/\la f |i\ra,
\ee
is the weak value \cite{aharonov1988result} of the system operator $A$, which has been prepared in state $|i\ra$ and postselected in state $|f\ra$, with probability $ p =|\la f|i\ra|^2$.  In the limit where the overlap between initial and final state goes to zero, the weak value can be arbitrarily large, but the probability then vanishes.  This effect can be seen as a signal amplification of the coupling constant $g$, a parameter we wish to measure.  We can quantify the precision of the measurement of $g$ with the signal-to-noise ratio (SNR), $\cal R$, defined as 
\be
{\cal R} \equiv \frac{ \la X_N\ra}{\sqrt{{\rm Var} [X_N]}},
\ee
where $X_N =(1/N) \sum_{k=1}^N x_k$ is the net meter signal acquired over $N$ events.  The signal mean is $\la X_N\ra$, once the $g=0$ background is subtracted, and ${\rm Var} [X_N]$ is the variance of the measured signal.  The meter is assumed to be in a Gaussian state characterized with width $\sigma$.  In the postselected meter state for standard weak-value amplification, the width of the meter remains the same, simply undergoing a shift in position.  Since we are considering a coherent state of light, the variance of the signal is given by the inverse number of photons detected, $N p$, times the squared width $\sigma^2$, while the signal is the shift $g A_w$ on the detector.  Consequently the SNR is given by ${\cal R} = g A_w \sqrt{p N}/\sigma = g \sqrt{N} \la f|A|i \ra/\sigma$.  The last factor, $\la f|A|i \ra$, can be made to be 1 for two-state systems, so the SNR is given by ${\cal R} = g \sqrt{N}/\sigma$ for a quantum-limited system. Note that this is the same SNR as without  weak-value amplification, as discussed in the introduction. 

The main effect we are investigating in this paper is that of recycling the events that would usually be thrown away.  By resetting the initial state, each round of recycling subtracts a fraction $p$ from the remaining light while keeping the signal and noise the same as above, but replacing $N$ with the remaining number of input photons.
Assuming $p\ll 1$, the amount of light remaining will decay exponentially with round number $j$ as $N_j = N (1-p)^j$, where $j =0,1,2,\ldots$.  The number of photons detected in round $j$ is then typically $p N_j$, resulting in exactly the same signal and noise as before, but with a new total photon number of $\sum_{j=0}^\infty p N_j = N$, in the limit of no losses.  Thus, the SNR ${\cal R} = g A_w \sqrt{N}/\sigma$ is itself weak-value amplified, giving an advantage both over the single-pass weak-value case as well as the direct standard method. 

{\it Experiment---}Our weak-measurement interferometer builds on the design of Dixon {\it et al.} \cite{dixon2009ultrasensitive}, whereby photons entering the Sagnac interferometer (lower right corner of Fig.~1) from the left experience nearly complete constructive interference of the CW and CCW paths, so that they are nearly all returned to that same port -- now an exit -- of the interferometer. 
In terms of (1) the initial state inside the Sagnac is $|i \ra = (e^{i \phi/2}|CW \ra + i e^{-i \phi/2}|CCW \ra)/\sqrt{2}$, where $\phi$ is the relative phase produced by a Soleil Babinet compensator.  The interferometer upper ``dark port" projects onto
$\la f| = (\la CW| + i \la CCW|)/\sqrt{2}$, with postselection probability $ |\la f|i\ra|^2 = \sin^2(\phi/2)$. A slight difference to the above theory discussion is that the weak value is imaginary, resulting in a shift in the complementary meter variable (the transverse spatial coordinate of the beam instead of its angle), but the above discussion is otherwise applicable. 

In the bright port we added a low-loss recycling loop with a polarization switch (implemented with a Pockels cell) that traps the pulses for up to 27 passes before another pulse is timed to enter the interferometer. A Position Sensitive Detector (PSD) designed for low-photon-number detection measures the beam displacement from tilting a mirror in a piezo-controlled mount, oscillated at a rate of 500 Hz (see Fig.~1), inside the Sagnac interferometer.  Thus, the parameter to be measured is the mirror tilt angle. We choose a beam width at the tilt mirror of 
$\sigma = 86 \mu$m, corresponding to an angular width of 0.94 mrad; the maximum applied angular mirror displacement is only 7.5 $\mu$rad, so the usual weak-value derivation approximations are well satisfied.

Our light source is a greatly attenuated 690-nm diode laser, pulsed at 200 kHz with approximately 600-ps pulsewidth; we attenuate the laser such that only one photon at a time is likely to be detected (i.e., our count rate on each detector is 20 kHz or less). Although of course weak-value measurements could also be made at the classical level, e.g., with a quad-cell photodetector and an unattenuated laser, we choose to work at the single-photon level so our system operates in a regime where the SNR is quantum-noise limited, i.e., dominated by photon shot noise. This allows for a direct test of the efficiency of recycling without the added complication of other systematic errors. 

The source photons are coupled through a single-mode fiber with a microscope objective and adjustable beam-expander at the fiber output, allowing us to control the input beam waist and position. A horizontal (H) polarizer is also placed at the fiber output. A 4f-imaging system prevents the beam from expanding as it traverses through the 1.2-meter loop, up to 27 times; a 300-cm focal length lens is placed at the optical center of both the recycling loop and the interferometer.  In order to maximize the weak-value SNR, a dark-port lens is added to image the surface of the tilting mirror (where the beam waist is located) onto the PSD; due to space constraints, a 50.2-mm focal length lens was placed 263 mm from the tilting mirror, and 62 mm from the PSD (resulting in a demagnified beam waist of $\sigma'$ = 20  $\mu$m). The PSD consists of a ``knife-edge'' mirror that divides the optical beam onto two Si avalanche photodiodes (APDs), with efficiencies \textasciitilde 65\% and dark counts \textless 250 cps; position information for the exiting photons is thus registered in the relative count rate differences between the two detectors.  Photon arrival timing information is recorded using a time tagger (quTools quTau) to make a cycle-by-cycle analysis. We measured that light hitting a \textasciitilde 3.75 $\mu$m-wide stripe at the very edge of the knife-edge mirror is scattered or absorbed, i.e., not detected by either APD; however, detailed modeling shows this PSD ``dead'' zone reduces the measured PSD signal by less than 0.5\%.

The number of round trips in the recycling system is limited by the postselection probability and the loss from scattering on optical components. We chose a postselection probability of $p=0.03$, corresponding to a relative phase $\phi = 0.35$ rad between the CW and CCW paths in the interferometer. Unlike the theoretical limits of this technique discussed above, loss from our optics is approximately 16$\%$, resulting in a total per-pass loss of about 19$\%$ including the postselection to the detector. The total loss was determined by fitting the intensity transmission versus pass number until less than half a percent of the light remained in the interferometer, showing an excellent exponential fit with the above loss-per-cycle.

{\it Results---}
We see approximately 3,000 counts per second in the first pass on each detector, and about 20,000 counts per second on each detector in all passes. Time tags are divided into 100-$\mu s$ bins (corresponding to 20 laser pulses or an average of about 2 photons per bin). The signal at the  500-Hz drive frequency of the piezo-driven mirror is extracted with a discrete Fourier transform of data taken over approximately 300 s, corresponding to a frequency resolution of 3.3 mHz. The noise, dominated by photon shot noise, is determined from the FFT, averaging over 100 frequencies  -- with 3.3-mHz spacings -- centered around the central mirror frequency. See Supplemental Material for more information.

\begin{figure}[tbh]
    \includegraphics[width=\linewidth]{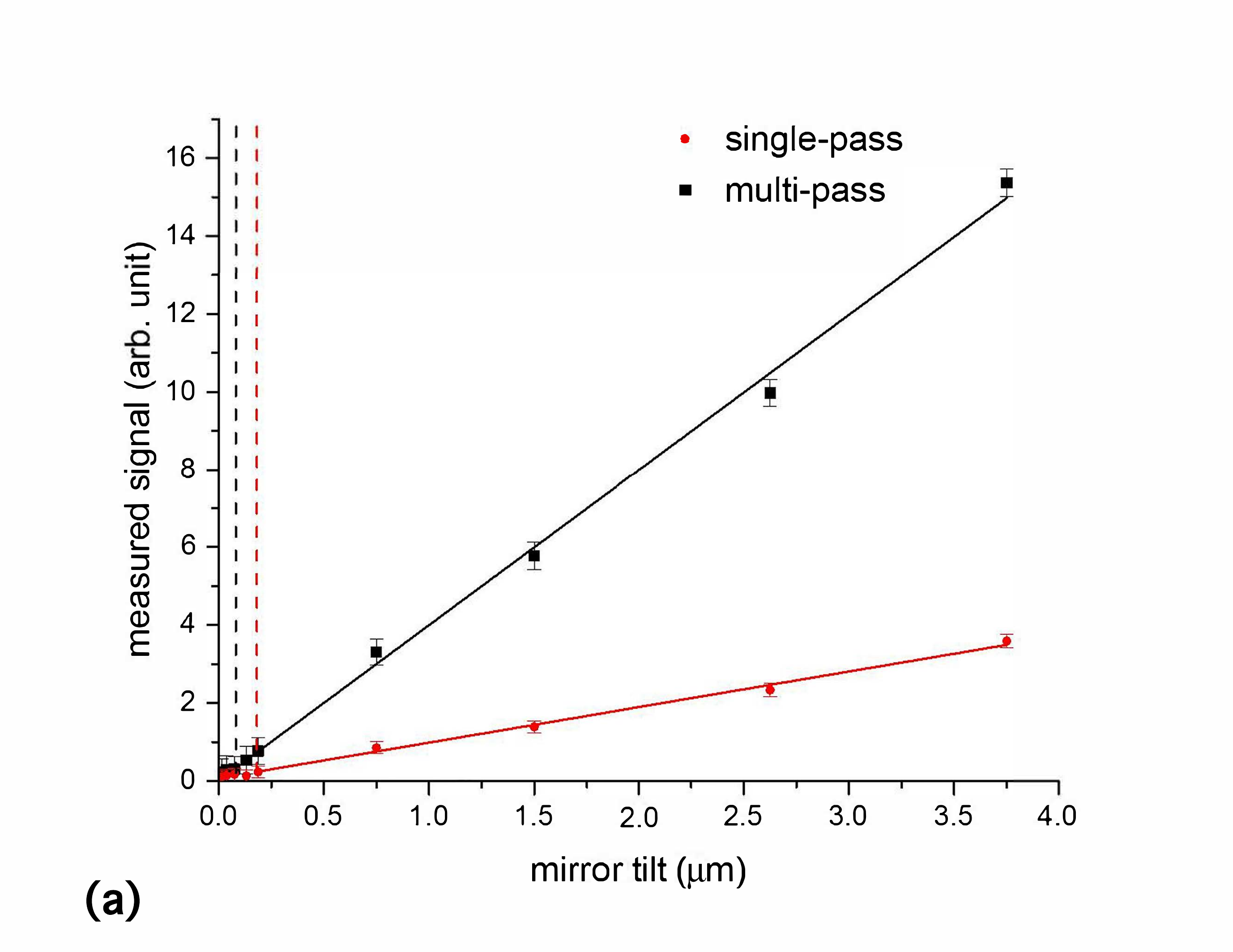}
    \includegraphics[width=\linewidth]{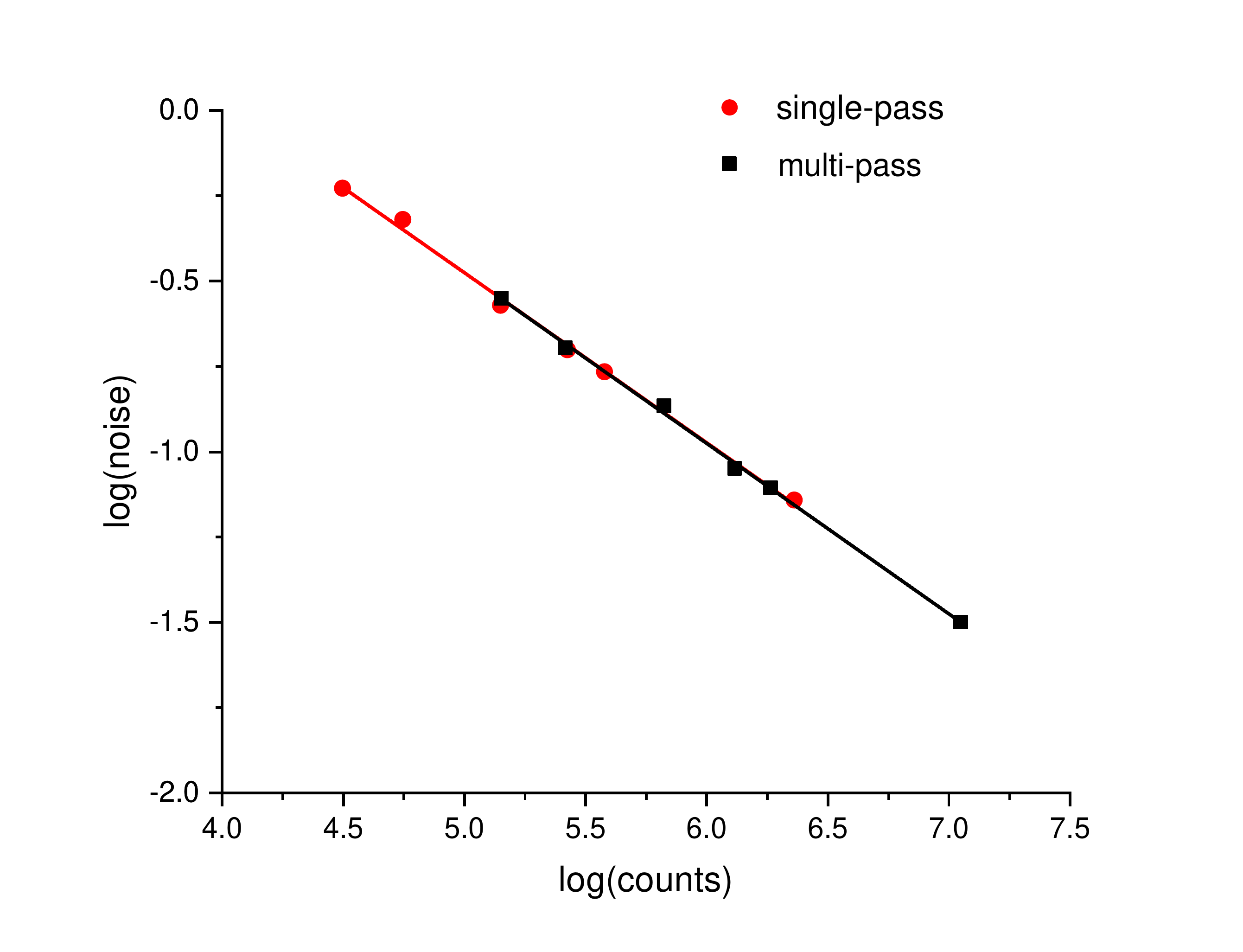}
    \caption{Data taken for single-pass weak-value amplification and multi-pass recycling weak-value amplification experiments. Panel (a) shows the signal increase from multiple recycling rounds.  Panel (b) shows that the noise decreases as the inverse square-root of the number of detected photons, indicating the experiment is shot-noise limited. The solid line shows the theoretical prediction for a $1/\sqrt{N}$ scaling of the relative noise, which appears linear on a log-log plot.}
    \label{fig:data1}
\end{figure}

\begin{figure}[tbh]
\includegraphics[width=\linewidth]{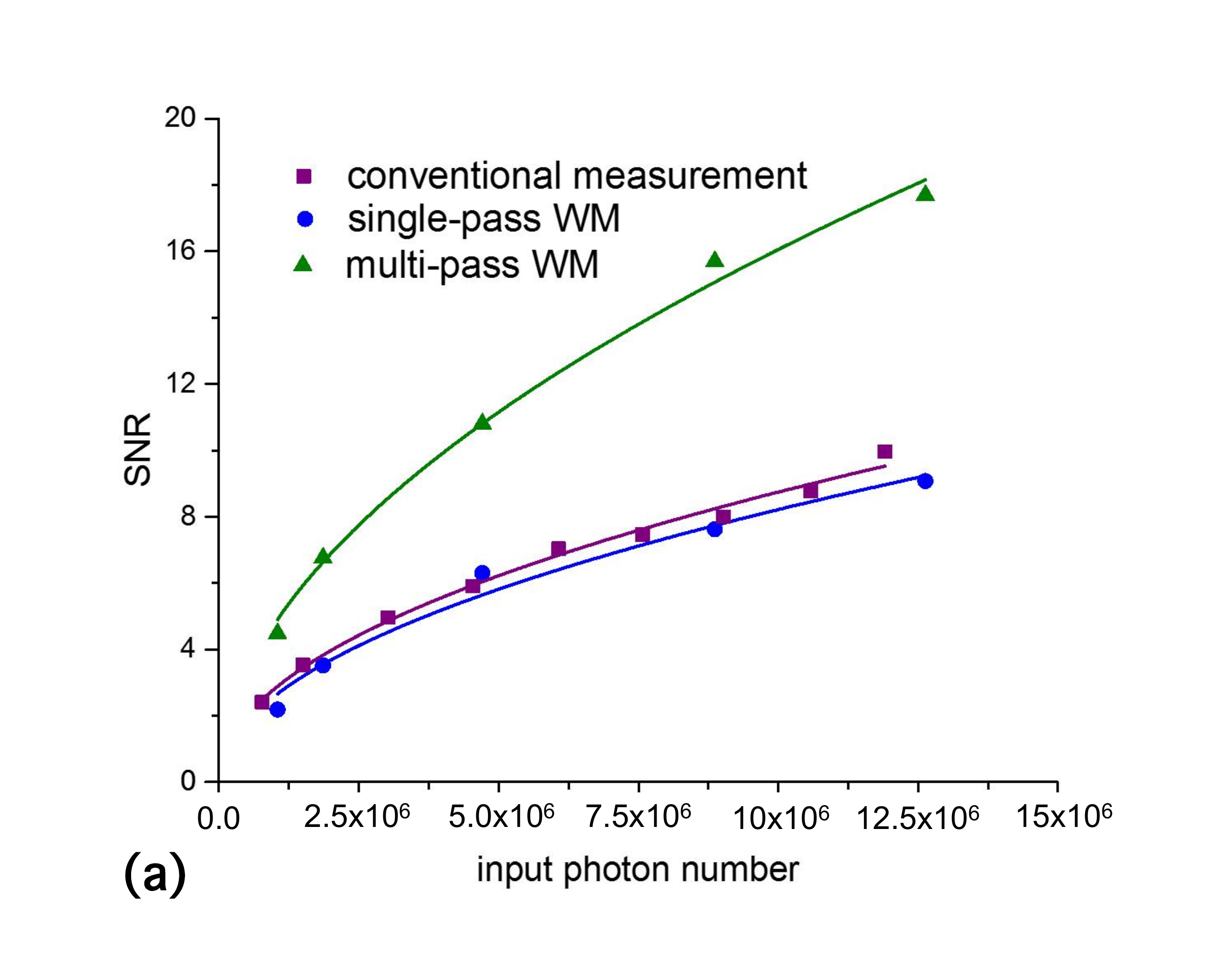}
    \includegraphics[width=\linewidth]{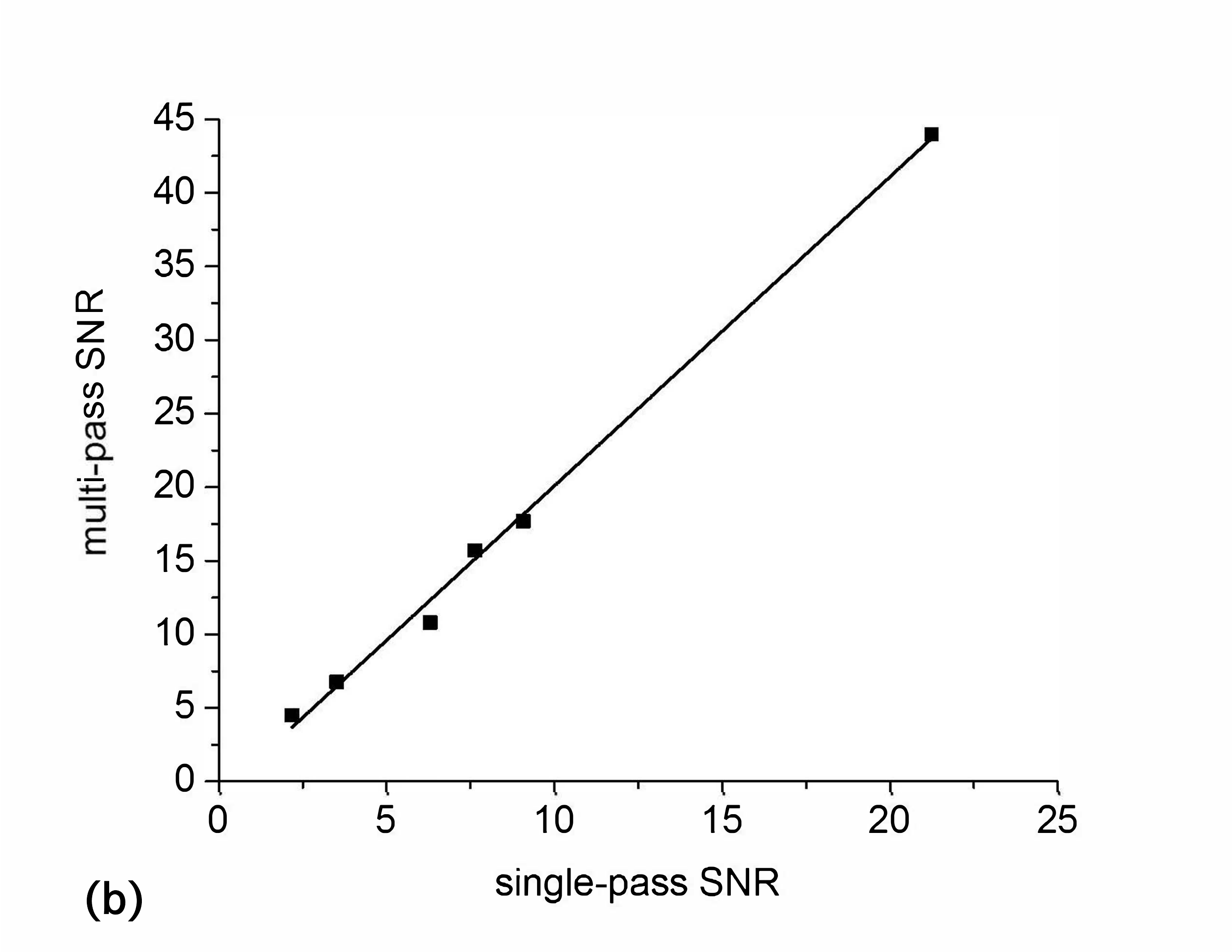}
    \caption{Data taken for conventional measurement, single-pass weak-value amplification, and multi-pass recycling weak-value amplification experiments. Panel (a) compares the SNR for the different experiments, showing scaling as the square-root of the photon number for all experiments.  Panel (b) plots the multi-pass experiment SNR vs. the single-pass experiment SNR, demonstrating improvement by a factor of $2.10 \pm 0.06$.}
    \label{fig:data2}
\end{figure}

Figs.~2-3 show our primary results. Fig.~2(a) shows the mean weak-value signal for both the single- and multi-pass cases. The average increase in counts with
recycling was 5 times that without recycling, so we expect our recycling signal to be 4.35 times the signal
without recycling, when beam reshaping effects are accounted for in our theoretical model; see Supplemental Information. We see a slope of $4.0 \pm 0.1 /\mu$rad ($0.91 \pm 0.04 /\mu$rad) for multi-pass (single-pass) measurements, representing a signal boost of $4.4 \pm 0.2$, in good agreement with our expectations.

Fig.~2(b) plots the noise floor, $\sqrt{{\rm Var}[X_N]}$, as a function of the number of detected photons.  As expected, this noise floor scales as $1/\sqrt{N}$, where $N$ is the number of detected photons, in both the single-and multiple-pass experiments; this verifies that the experiment is operating in the photon shot-noise limit. Note that although the noise scaling is the same for the single- and multi-pass configurations, the actual noise will be $\sim \sqrt{5}$ times higher in the latter, due to the higher number of counts.

In order to properly evaluate the relative sensitivity of the weak-value measurement, we compare it directly with the results of a conventional measurement of the mirror tilt, using the same wavelength, number of photons and beam waist on the mirror (the only factors to affect the conventional measurement SNR). To do this, we effectively blocked one path through the interferometer by placing a polarizer between the 50/50 beamsplitter and the half-wave plate (see Fig. 1). The dark-port imaging lens was replaced with an  {\it f} = 300 mm focal-length lens placed 300 mm in front of the PSD, so that mirror tilts were converted to lateral displacements at the PSD, e.g., our maximum 7.5-$\mu$rad tilt angle produced a shift of 2.25 $\mu$m on the 280-$\mu$m beam waist at the PSD; 
note that since both the waist size and displacement at the PSD are proportional to {\it f}, the SNR is independent of {\it f}. We then applied the same data analysis procedure as with the weak measurements.

Fig.~3(a) plots the SNR, ${\cal R}$, for three different experiments:  The conventional measurement of the mirror tilt, using all the input photons, is plotted in purple squares, while the single-pass (multi-pass) weak-value measurement is plotted using blue circles (green triangles). We can clearly see that all scale as $\sqrt{N}$, where $N$ is the input photon number, as expected for a quantum-limited experiment \cite{starling2009optimizing}. However, the multi-pass data displays substantially larger SNR than the single-pass weak-value case and the conventional measurement, which have nearly the same SNR, as expected.   

Fig.~3(b) plots the SNR of the multi-pass experiment versus the SNR of the single-pass experiment.  A good linear fit is observed, with a fitted boost $2.10 \pm 0.06$, to be compared with $ \sqrt{N_{all\:  passes}/N_{1\:pass}} = \sqrt{5}= 2.4$. Our measured value is slightly lowered due to a predicted beam-reshaping effect (see Supplemental Information); the resulting modified theoretical prediction is 1.95, in reasonable agreement with our measurement.

{\it Discussion and Outlook---} Using photon recycled weak-value amplification, we have demonstrated a $>2\times$ boost in the SNR of an optical mirror tilt measurement compared to conventional, quantum-limited measurement.  Our method scavenges the wasted photons of a standard weak-value amplification measurement in order to use them as a resource.  Our current experiments were limited by loss in the optical elements of 16$\%$, with a postselection fraction of $3\%$ per cycle.  If the optical loss were made lower than the postselection fraction, this experiment would have a SNR boost of about 6.  

In our current implementation the photons are only in the system for up to $\sim$100 ns, but the existing switch-in Pockel cell could only be fired every 5 $\mu$s, due to high-voltage driver limitations, limiting our duty cycle to 2\%.
However, there are other switch methods (e.g., relying on nonlinear optics \cite{Kupchak2019}) which could enable switch repetition rates exceeding 10 GHz, and switching times below 1 ps (thereby allowing a much shorter recycling loop, assuming the loss in these switches could be minimized); thus, one can envision a system operating at rates some four orders of magnitude beyond our current demonstration. Note that
the technique we presented here is a kind of power recycling (see e.g. \cite{zhang2018simultaneously}), based on discrete pulse trapping (see also recent theory work \cite{fang2020ultrasensitive}); a complementary method of doing the same thing uses passive optical cavities to boost the number of photons undergoing weak-value amplification \cite{lyons2015power,wang2016experimental,fang2019ultrasensitive}. Our proof-of-principle experiment shows a promising way forward to further improve on the already excellent performance of weak-value amplification experiments.

{\it Acknowledgements}---We thank Kevin Lyons for helpful discussions, and Spencer Johnson for precise measurements of our PSD knife-edge mirror. This work was supported by the U.S. Army Research Office (ARO), under Award No. W911NF-13-1-0402. ANJ also acknowledges support from NSF grant no. DMR-1809343.

\section{Supplemental Material}
{\it Data analysis}---
After collecting and binning the raw detector counts for 300 s at a given mirror tilt amplitude, the difference between the counts recorded by the left and right detectors is taken for each 100-$\mu s$ time bin. We extract the signal from the FFT at the 50-Hz oscillation frequency, take the absolute value, normalize by the total number of counts from the entire 300 s, multiply by $\sqrt{\pi/2}$ (a normalization factor coming from the fact that our beam is Gaussian), and multiply by 4 to convert the FFT output to the peak-to-peak
amplitude. This process is repeated 20 times for each data point.

In order to simultaneously gather many samples to compute the noise, we use the root mean square of the FFT signal from 100 frequencies spaced 3 mHz from the oscillation frequency, again normalized and scaled. It is possible that by including only off-signal frequencies in our analysis, we miss some noise occurring
only at the signal frequency, which may happen, for example, if the mirror or piezo itself has some extra
uncertainty in the oscillation amplitude. To confirm that this was not the case, we took 100 repeated
measurements with a 3-$\mu$rad peak-to-peak mirror tilt amplitude; the standard deviation of those measurements was consistent with the expected standard deviation from shot noise, confirming that it is the dominant effect.

{\it Beam reshaping}---
In the original interferometric weak value scheme, the detector collected a power of $P_d= p P$, where $p$ was the post-selected fraction of the total laser power $P$ coming from the dark port. If we recycle the unused light, however, the average power $P_d$ collected at the detector after $r$ recycling passes 
will have the modified form 
\begin{align}\label{eq:power}
  P_d &= \sum_{n=1}^{r} (1-p)^{n-1} p P = (1-(1-p)^{r}) P,
\end{align}
where $p$ is the fraction of the input power that exits the dark port of the interferometer after each traversal, and $r$ is the number of recycled pulses that hit the detector.  Here we have ignored optical losses and detector inefficiencies for clarity.  The power collected at the detector after a single traversal is $p P$ and the SNR scales as $\sqrt{P_d}$, so the net SNR gain factor will be
\begin{align}\label{eq:powergain}
  \sqrt{\frac{P_d}{p P}} &= \sqrt{\frac{1 - (1-p)^{r}}{p}}.
\end{align}

For a small post-selection probability---such as those used in weak-value experiments--- we can then expand \eqref{eq:powergain} around $p = 0$ to find
\begin{align}
  \sqrt{\frac{P_d}{p P}} &\approx \sqrt{r}\left(1 - (r-1) \frac{p}{4}\right) + O(p^2).
\end{align}
For $p (r-1) \ll 1$, we can neglect the attenuation of the pulse to see an approximate $\sqrt{r}$ SNR scaling \cite{dressel2013strengthening}.

A more detailed analysis of the transverse beam profile can be carried out in the multi-cycle case. Given a mirror momentum kick $k$, a relative phase shift $\phi$ between the paths, and a loss per cycle $\gamma$, the number density $n_r(x)$ of photons that hit the dark port detector at a transverse position $x$ on the $r$\textsuperscript{th} pulse traversal is,
\begin{align}\label{eq:nminus}
  n_r(x) &= n_0(x) (1-\gamma)^r \times \\
  &\qquad \sin^2\left(\frac{\phi}{2} - kx\right)\cos^{2(r-1)}\left(\frac{\phi}{2} - kx\right), \nonumber
\end{align}
where $n_0(x)$ is the number density for the input pulse, well-approximated as a Gaussian distribution \cite{dressel2013strengthening}.

The total number density $\bar{n}_r(x)$ that accumulates on the dark port detector after $r$ traversals of the pulse will be the sum of the number densities for the $r$ traversals,
\begin{align}\label{eq:nminusacc}
  \bar{n}_r(x) &= \sum_{j=1}^r n_j(x) \\
  &= n_0(x) \frac{(1-\gamma)\left(1 - \left[(1-\gamma) \cos^{2}\left(\phi/2 - kx\right)\right]^r\right)}{1 + \gamma \cot^2\left(\phi/2 - kx\right)}. \nonumber
\end{align}
Hence, the total number of photons that hit the detector after $r$ traversals is $N_r = \int \! \textrm{d}x\, \bar{n}_r(x)$. 

In the limit of an infinite number of trials $r \to \infty$, the final term (in []'s) in \eqref{eq:nminusacc} vanishes and we are left with the number density,
\begin{align}\label{eq:nminusacclimit}
  \bar{n}_\infty(x) &= n_0(x) \frac{1-\gamma}{1 + \gamma \cot^2\left(\phi/2 - kx\right)}.
\end{align}
For no loss, $\gamma \to 0$, the modulating factor from the measurement cancels and the original pulse shape is perfectly recovered!  The reason for this is that although the initial photons are weak-value amplified, this effect results in back-action on the remaining photons in the beam.  The selection of the weak value-amplified events ``etches'' the profile, so the distribution walks-off.  With no loss, the net result will be no signal at all \cite{dressel2013strengthening}!
However, because our photons experience an additional 16$\%$ loss from the optics in our setup, the beam reshaping effect is greatly reduced. 

Note that an alternative method to greatly reduce this deleterious effect -- even in the low-loss case -- would be to arrange for the post-selected distribution to be pulled from opposite sides of the beam profile with each subsequent pass, which could be accomplished by designing an {\it odd} number of reflections into the system, such that the profile is flipped on consecutive passes.  This simple fix causes the integrated profile (in the $\gamma \rightarrow 0$ limit) to be
\begin{align}
  \bar{n}_{\infty}(x) &= n_0(x) \frac{\sin^2\left(\frac{\phi}{2} -kx\right)\left(1+\cos^2\left(\frac{\phi}{2} - kx\right)\right)}{1 - \cos^2\left(\frac{\phi}{2} - kx\right)\cos^2\left(\frac{\phi}{2} + kx\right)}.
\end{align} 
For small $k \sigma$ and $\phi$, this result recovers the full weak-value amplification effect, for all the photons \cite{dressel2013strengthening}.

\bibliography{refs}

\begin{thebibliography}{30}%
\makeatletter
\providecommand \@ifxundefined [1]{%
 \@ifx{#1\undefined}
}%
\providecommand \@ifnum [1]{%
 \ifnum #1\expandafter \@firstoftwo
 \else \expandafter \@secondoftwo
 \fi
}%
\providecommand \@ifx [1]{%
 \ifx #1\expandafter \@firstoftwo
 \else \expandafter \@secondoftwo
 \fi
}%
\providecommand \natexlab [1]{#1}%
\providecommand \enquote  [1]{``#1''}%
\providecommand \bibnamefont  [1]{#1}%
\providecommand \bibfnamefont [1]{#1}%
\providecommand \citenamefont [1]{#1}%
\providecommand \href@noop [0]{\@secondoftwo}%
\providecommand \href [0]{\begingroup \@sanitize@url \@href}%
\providecommand \@href[1]{\@@startlink{#1}\@@href}%
\providecommand \@@href[1]{\endgroup#1\@@endlink}%
\providecommand \@sanitize@url [0]{\catcode `\\12\catcode `\$12\catcode
  `\&12\catcode `\#12\catcode `\^12\catcode `\_12\catcode `\%12\relax}%
\providecommand \@@startlink[1]{}%
\providecommand \@@endlink[0]{}%
\providecommand \url  [0]{\begingroup\@sanitize@url \@url }%
\providecommand \@url [1]{\endgroup\@href {#1}{\urlprefix }}%
\providecommand \urlprefix  [0]{URL }%
\providecommand \Eprint [0]{\href }%
\providecommand \doibase [0]{http://dx.doi.org/}%
\providecommand \selectlanguage [0]{\@gobble}%
\providecommand \bibinfo  [0]{\@secondoftwo}%
\providecommand \bibfield  [0]{\@secondoftwo}%
\providecommand \translation [1]{[#1]}%
\providecommand \BibitemOpen [0]{}%
\providecommand \bibitemStop [0]{}%
\providecommand \bibitemNoStop [0]{.\EOS\space}%
\providecommand \EOS [0]{\spacefactor3000\relax}%
\providecommand \BibitemShut  [1]{\csname bibitem#1\endcsname}%
\let\auto@bib@innerbib\@empty
\bibitem [{\citenamefont {Aharonov}\ \emph {et~al.}(1988)\citenamefont
  {Aharonov}, \citenamefont {Albert},\ and\ \citenamefont
  {Vaidman}}]{aharonov1988result}%
  \BibitemOpen
  \bibfield  {author} {\bibinfo {author} {\bibfnamefont {Y.}~\bibnamefont
  {Aharonov}}, \bibinfo {author} {\bibfnamefont {D.~Z.}\ \bibnamefont
  {Albert}}, \ and\ \bibinfo {author} {\bibfnamefont {L.}~\bibnamefont
  {Vaidman}},\ }\href@noop {} {\bibfield  {journal} {\bibinfo  {journal} {Phys.
  Rev. Lett.}\ }\textbf {\bibinfo {volume} {60}},\ \bibinfo {pages} {1351}
  (\bibinfo {year} {1988})}\BibitemShut {NoStop}%
\bibitem [{\citenamefont {Dressel}\ \emph {et~al.}(2014)\citenamefont
  {Dressel}, \citenamefont {Malik}, \citenamefont {Miatto}, \citenamefont
  {Jordan},\ and\ \citenamefont {Boyd}}]{dressel2014colloquium}%
  \BibitemOpen
  \bibfield  {author} {\bibinfo {author} {\bibfnamefont {J.}~\bibnamefont
  {Dressel}}, \bibinfo {author} {\bibfnamefont {M.}~\bibnamefont {Malik}},
  \bibinfo {author} {\bibfnamefont {F.~M.}\ \bibnamefont {Miatto}}, \bibinfo
  {author} {\bibfnamefont {A.~N.}\ \bibnamefont {Jordan}}, \ and\ \bibinfo
  {author} {\bibfnamefont {R.~W.}\ \bibnamefont {Boyd}},\ }\href@noop {}
  {\bibfield  {journal} {\bibinfo  {journal} {Rev. Mod. Phys.}\ }\textbf
  {\bibinfo {volume} {86}},\ \bibinfo {pages} {307} (\bibinfo {year}
  {2014})}\BibitemShut {NoStop}%
\bibitem [{\citenamefont {Hosten}\ and\ \citenamefont
  {Kwiat}(2008)}]{hosten2008observation}%
  \BibitemOpen
  \bibfield  {author} {\bibinfo {author} {\bibfnamefont {O.}~\bibnamefont
  {Hosten}}\ and\ \bibinfo {author} {\bibfnamefont {P.}~\bibnamefont {Kwiat}},\
  }\href@noop {} {\bibfield  {journal} {\bibinfo  {journal} {Science}\ }\textbf
  {\bibinfo {volume} {319}},\ \bibinfo {pages} {787} (\bibinfo {year}
  {2008})}\BibitemShut {NoStop}%
\bibitem [{\citenamefont {Dixon}\ \emph {et~al.}(2009)\citenamefont {Dixon},
  \citenamefont {Starling}, \citenamefont {Jordan},\ and\ \citenamefont
  {Howell}}]{dixon2009ultrasensitive}%
  \BibitemOpen
  \bibfield  {author} {\bibinfo {author} {\bibfnamefont {P.~B.}\ \bibnamefont
  {Dixon}}, \bibinfo {author} {\bibfnamefont {D.~J.}\ \bibnamefont {Starling}},
  \bibinfo {author} {\bibfnamefont {A.~N.}\ \bibnamefont {Jordan}}, \ and\
  \bibinfo {author} {\bibfnamefont {J.~C.}\ \bibnamefont {Howell}},\
  }\href@noop {} {\bibfield  {journal} {\bibinfo  {journal} {Phys. Rev. Lett.}\
  }\textbf {\bibinfo {volume} {102}},\ \bibinfo {pages} {173601} (\bibinfo
  {year} {2009})}\BibitemShut {NoStop}%
\bibitem [{\citenamefont {Starling}\ \emph {et~al.}(2010)\citenamefont
  {Starling}, \citenamefont {Dixon}, \citenamefont {Jordan},\ and\
  \citenamefont {Howell}}]{starling2010precision}%
  \BibitemOpen
  \bibfield  {author} {\bibinfo {author} {\bibfnamefont {D.~J.}\ \bibnamefont
  {Starling}}, \bibinfo {author} {\bibfnamefont {P.~B.}\ \bibnamefont {Dixon}},
  \bibinfo {author} {\bibfnamefont {A.~N.}\ \bibnamefont {Jordan}}, \ and\
  \bibinfo {author} {\bibfnamefont {J.~C.}\ \bibnamefont {Howell}},\
  }\href@noop {} {\bibfield  {journal} {\bibinfo  {journal} {Phys. Rev. A}\
  }\textbf {\bibinfo {volume} {82}},\ \bibinfo {pages} {063822} (\bibinfo
  {year} {2010})}\BibitemShut {NoStop}%
\bibitem [{\citenamefont {Viza}\ \emph {et~al.}(2013)\citenamefont {Viza},
  \citenamefont {Mart{\'\i}nez-Rinc{\'o}n}, \citenamefont {Howland},
  \citenamefont {Frostig}, \citenamefont {Shomroni}, \citenamefont {Dayan},\
  and\ \citenamefont {Howell}}]{viza2013weak}%
  \BibitemOpen
  \bibfield  {author} {\bibinfo {author} {\bibfnamefont {G.~I.}\ \bibnamefont
  {Viza}}, \bibinfo {author} {\bibfnamefont {J.}~\bibnamefont
  {Mart{\'\i}nez-Rinc{\'o}n}}, \bibinfo {author} {\bibfnamefont {G.~A.}\
  \bibnamefont {Howland}}, \bibinfo {author} {\bibfnamefont {H.}~\bibnamefont
  {Frostig}}, \bibinfo {author} {\bibfnamefont {I.}~\bibnamefont {Shomroni}},
  \bibinfo {author} {\bibfnamefont {B.}~\bibnamefont {Dayan}}, \ and\ \bibinfo
  {author} {\bibfnamefont {J.~C.}\ \bibnamefont {Howell}},\ }\href@noop {}
  {\bibfield  {journal} {\bibinfo  {journal} {Opt. Lett.}\ }\textbf {\bibinfo
  {volume} {38}},\ \bibinfo {pages} {2949} (\bibinfo {year}
  {2013})}\BibitemShut {NoStop}%
\bibitem [{\citenamefont {Egan}\ and\ \citenamefont
  {Stone}(2012)}]{egan2012weak}%
  \BibitemOpen
  \bibfield  {author} {\bibinfo {author} {\bibfnamefont {P.}~\bibnamefont
  {Egan}}\ and\ \bibinfo {author} {\bibfnamefont {J.~A.}\ \bibnamefont
  {Stone}},\ }\href@noop {} {\bibfield  {journal} {\bibinfo  {journal} {Opt.
  Lett.}\ }\textbf {\bibinfo {volume} {37}},\ \bibinfo {pages} {4991} (\bibinfo
  {year} {2012})}\BibitemShut {NoStop}%
\bibitem [{\citenamefont {Li}\ \emph {et~al.}(2016)\citenamefont {Li},
  \citenamefont {Shen}, \citenamefont {He}, \citenamefont {Zhang},
  \citenamefont {Chen},\ and\ \citenamefont {Ma}}]{li2016application}%
  \BibitemOpen
  \bibfield  {author} {\bibinfo {author} {\bibfnamefont {D.}~\bibnamefont
  {Li}}, \bibinfo {author} {\bibfnamefont {Z.}~\bibnamefont {Shen}}, \bibinfo
  {author} {\bibfnamefont {Y.}~\bibnamefont {He}}, \bibinfo {author}
  {\bibfnamefont {Y.}~\bibnamefont {Zhang}}, \bibinfo {author} {\bibfnamefont
  {Z.}~\bibnamefont {Chen}}, \ and\ \bibinfo {author} {\bibfnamefont
  {H.}~\bibnamefont {Ma}},\ }\href@noop {} {\bibfield  {journal} {\bibinfo
  {journal} {Appl. Opt.}\ }\textbf {\bibinfo {volume} {55}},\ \bibinfo {pages}
  {1697} (\bibinfo {year} {2016})}\BibitemShut {NoStop}%
\bibitem [{\citenamefont {Qu}\ \emph {et~al.}(2018)\citenamefont {Qu},
  \citenamefont {Sun}, \citenamefont {Jin}, \citenamefont {Jiang},
  \citenamefont {Wen},\ and\ \citenamefont {Xiao}}]{qu2018sub}%
  \BibitemOpen
  \bibfield  {author} {\bibinfo {author} {\bibfnamefont {W.}~\bibnamefont
  {Qu}}, \bibinfo {author} {\bibfnamefont {J.}~\bibnamefont {Sun}}, \bibinfo
  {author} {\bibfnamefont {S.}~\bibnamefont {Jin}}, \bibinfo {author}
  {\bibfnamefont {L.}~\bibnamefont {Jiang}}, \bibinfo {author} {\bibfnamefont
  {J.}~\bibnamefont {Wen}}, \ and\ \bibinfo {author} {\bibfnamefont
  {Y.}~\bibnamefont {Xiao}},\ }\href@noop {} {\bibfield  {journal} {\bibinfo
  {journal} {arXiv:1812.02551}\ } (\bibinfo {year} {2018})}\BibitemShut
  {NoStop}%
\bibitem [{\citenamefont {Vella}\ \emph {et~al.}(2019)\citenamefont {Vella},
  \citenamefont {Head}, \citenamefont {Brown},\ and\ \citenamefont
  {Alonso}}]{vella2019simultaneous}%
  \BibitemOpen
  \bibfield  {author} {\bibinfo {author} {\bibfnamefont {A.}~\bibnamefont
  {Vella}}, \bibinfo {author} {\bibfnamefont {S.~T.}\ \bibnamefont {Head}},
  \bibinfo {author} {\bibfnamefont {T.~G.}\ \bibnamefont {Brown}}, \ and\
  \bibinfo {author} {\bibfnamefont {M.~A.}\ \bibnamefont {Alonso}},\
  }\href@noop {} {\bibfield  {journal} {\bibinfo  {journal} {Phys. Rev. Lett.}\
  }\textbf {\bibinfo {volume} {122}},\ \bibinfo {pages} {123603} (\bibinfo
  {year} {2019})}\BibitemShut {NoStop}%
\bibitem [{\citenamefont {Salazar-Serrano}\ \emph {et~al.}(2015)\citenamefont
  {Salazar-Serrano}, \citenamefont {Guzm{\'a}n}, \citenamefont {Valencia},\
  and\ \citenamefont {Torres}}]{salazar2015demonstration}%
  \BibitemOpen
  \bibfield  {author} {\bibinfo {author} {\bibfnamefont {L.~J.}\ \bibnamefont
  {Salazar-Serrano}}, \bibinfo {author} {\bibfnamefont {D.~A.}\ \bibnamefont
  {Guzm{\'a}n}}, \bibinfo {author} {\bibfnamefont {A.}~\bibnamefont
  {Valencia}}, \ and\ \bibinfo {author} {\bibfnamefont {J.~P.}\ \bibnamefont
  {Torres}},\ }\href@noop {} {\bibfield  {journal} {\bibinfo  {journal} {Opt.
  express}\ }\textbf {\bibinfo {volume} {23}},\ \bibinfo {pages} {10097}
  (\bibinfo {year} {2015})}\BibitemShut {NoStop}%
\bibitem [{\citenamefont {Starling}\ \emph {et~al.}(2009)\citenamefont
  {Starling}, \citenamefont {Dixon}, \citenamefont {Jordan},\ and\
  \citenamefont {Howell}}]{starling2009optimizing}%
  \BibitemOpen
  \bibfield  {author} {\bibinfo {author} {\bibfnamefont {D.~J.}\ \bibnamefont
  {Starling}}, \bibinfo {author} {\bibfnamefont {P.~B.}\ \bibnamefont {Dixon}},
  \bibinfo {author} {\bibfnamefont {A.~N.}\ \bibnamefont {Jordan}}, \ and\
  \bibinfo {author} {\bibfnamefont {J.~C.}\ \bibnamefont {Howell}},\
  }\href@noop {} {\bibfield  {journal} {\bibinfo  {journal} {Phys. Rev. A}\
  }\textbf {\bibinfo {volume} {80}},\ \bibinfo {pages} {041803} (\bibinfo
  {year} {2009})}\BibitemShut {NoStop}%
\bibitem [{\citenamefont {Feizpour}\ \emph {et~al.}(2011)\citenamefont
  {Feizpour}, \citenamefont {Xing},\ and\ \citenamefont
  {Steinberg}}]{feizpour2011amplifying}%
  \BibitemOpen
  \bibfield  {author} {\bibinfo {author} {\bibfnamefont {A.}~\bibnamefont
  {Feizpour}}, \bibinfo {author} {\bibfnamefont {X.}~\bibnamefont {Xing}}, \
  and\ \bibinfo {author} {\bibfnamefont {A.~M.}\ \bibnamefont {Steinberg}},\
  }\href@noop {} {\bibfield  {journal} {\bibinfo  {journal} {Phys. Rev. Lett.}\
  }\textbf {\bibinfo {volume} {107}},\ \bibinfo {pages} {133603} (\bibinfo
  {year} {2011})}\BibitemShut {NoStop}%
\bibitem [{\citenamefont {Sinclair}\ \emph {et~al.}(2017)\citenamefont
  {Sinclair}, \citenamefont {Hallaji}, \citenamefont {Steinberg}, \citenamefont
  {Tollaksen},\ and\ \citenamefont {Jordan}}]{sinclair2017weak}%
  \BibitemOpen
  \bibfield  {author} {\bibinfo {author} {\bibfnamefont {J.}~\bibnamefont
  {Sinclair}}, \bibinfo {author} {\bibfnamefont {M.}~\bibnamefont {Hallaji}},
  \bibinfo {author} {\bibfnamefont {A.~M.}\ \bibnamefont {Steinberg}}, \bibinfo
  {author} {\bibfnamefont {J.}~\bibnamefont {Tollaksen}}, \ and\ \bibinfo
  {author} {\bibfnamefont {A.~N.}\ \bibnamefont {Jordan}},\ }\href@noop {}
  {\bibfield  {journal} {\bibinfo  {journal} {Phys. Rev. A}\ }\textbf {\bibinfo
  {volume} {96}},\ \bibinfo {pages} {052128} (\bibinfo {year}
  {2017})}\BibitemShut {NoStop}%
\bibitem [{\citenamefont {Pang}\ \emph {et~al.}(2016)\citenamefont {Pang},
  \citenamefont {Alonso}, \citenamefont {Brun},\ and\ \citenamefont
  {Jordan}}]{pang2016protecting}%
  \BibitemOpen
  \bibfield  {author} {\bibinfo {author} {\bibfnamefont {S.}~\bibnamefont
  {Pang}}, \bibinfo {author} {\bibfnamefont {J.~R.~G.}\ \bibnamefont {Alonso}},
  \bibinfo {author} {\bibfnamefont {T.~A.}\ \bibnamefont {Brun}}, \ and\
  \bibinfo {author} {\bibfnamefont {A.~N.}\ \bibnamefont {Jordan}},\
  }\href@noop {} {\bibfield  {journal} {\bibinfo  {journal} {Phys. Rev. A}\
  }\textbf {\bibinfo {volume} {94}},\ \bibinfo {pages} {012329} (\bibinfo
  {year} {2016})}\BibitemShut {NoStop}%
\bibitem [{\citenamefont {Li}\ \emph {et~al.}(2018)\citenamefont {Li},
  \citenamefont {Li}, \citenamefont {Zhang}, \citenamefont {Yu}, \citenamefont
  {Lu}, \citenamefont {Liu}, \citenamefont {Zhang},\ and\ \citenamefont
  {Pan}}]{li2018phase}%
  \BibitemOpen
  \bibfield  {author} {\bibinfo {author} {\bibfnamefont {L.}~\bibnamefont
  {Li}}, \bibinfo {author} {\bibfnamefont {Y.}~\bibnamefont {Li}}, \bibinfo
  {author} {\bibfnamefont {Y.-L.}\ \bibnamefont {Zhang}}, \bibinfo {author}
  {\bibfnamefont {S.}~\bibnamefont {Yu}}, \bibinfo {author} {\bibfnamefont
  {C.-Y.}\ \bibnamefont {Lu}}, \bibinfo {author} {\bibfnamefont {N.-L.}\
  \bibnamefont {Liu}}, \bibinfo {author} {\bibfnamefont {J.}~\bibnamefont
  {Zhang}}, \ and\ \bibinfo {author} {\bibfnamefont {J.-W.}\ \bibnamefont
  {Pan}},\ }\href@noop {} {\bibfield  {journal} {\bibinfo  {journal} {Phys.
  Rev. A}\ }\textbf {\bibinfo {volume} {97}},\ \bibinfo {pages} {033851}
  (\bibinfo {year} {2018})}\BibitemShut {NoStop}%
\bibitem [{\citenamefont {Viza}\ \emph {et~al.}(2015)\citenamefont {Viza},
  \citenamefont {Mart{\'\i}nez-Rinc{\'o}n}, \citenamefont {Alves},
  \citenamefont {Jordan},\ and\ \citenamefont
  {Howell}}]{viza2015experimentally}%
  \BibitemOpen
  \bibfield  {author} {\bibinfo {author} {\bibfnamefont {G.~I.}\ \bibnamefont
  {Viza}}, \bibinfo {author} {\bibfnamefont {J.}~\bibnamefont
  {Mart{\'\i}nez-Rinc{\'o}n}}, \bibinfo {author} {\bibfnamefont {G.~B.}\
  \bibnamefont {Alves}}, \bibinfo {author} {\bibfnamefont {A.~N.}\ \bibnamefont
  {Jordan}}, \ and\ \bibinfo {author} {\bibfnamefont {J.~C.}\ \bibnamefont
  {Howell}},\ }\href@noop {} {\bibfield  {journal} {\bibinfo  {journal} {Phys.
  Rev. A}\ }\textbf {\bibinfo {volume} {92}},\ \bibinfo {pages} {032127}
  (\bibinfo {year} {2015})}\BibitemShut {NoStop}%
\bibitem [{\citenamefont {Jordan}\ \emph {et~al.}(2014)\citenamefont {Jordan},
  \citenamefont {Mart{\'\i}nez-Rinc{\'o}n},\ and\ \citenamefont
  {Howell}}]{jordan2014technical}%
  \BibitemOpen
  \bibfield  {author} {\bibinfo {author} {\bibfnamefont {A.~N.}\ \bibnamefont
  {Jordan}}, \bibinfo {author} {\bibfnamefont {J.}~\bibnamefont
  {Mart{\'\i}nez-Rinc{\'o}n}}, \ and\ \bibinfo {author} {\bibfnamefont {J.~C.}\
  \bibnamefont {Howell}},\ }\href@noop {} {\bibfield  {journal} {\bibinfo
  {journal} {Phys. Rev. X}\ }\textbf {\bibinfo {volume} {4}},\ \bibinfo {pages}
  {011031} (\bibinfo {year} {2014})}\BibitemShut {NoStop}%
\bibitem [{\citenamefont {Lyons}\ \emph {et~al.}(2018)\citenamefont {Lyons},
  \citenamefont {Howell},\ and\ \citenamefont {Jordan}}]{lyons2018noise}%
  \BibitemOpen
  \bibfield  {author} {\bibinfo {author} {\bibfnamefont {K.}~\bibnamefont
  {Lyons}}, \bibinfo {author} {\bibfnamefont {J.~C.}\ \bibnamefont {Howell}}, \
  and\ \bibinfo {author} {\bibfnamefont {A.~N.}\ \bibnamefont {Jordan}},\
  }\href@noop {} {\bibfield  {journal} {\bibinfo  {journal} {Quantum Stud.:
  Math. Found.}\ }\textbf {\bibinfo {volume} {5}},\ \bibinfo {pages} {579}
  (\bibinfo {year} {2018})}\BibitemShut {NoStop}%
\bibitem [{\citenamefont {Dressel}\ \emph {et~al.}(2013)\citenamefont
  {Dressel}, \citenamefont {Lyons}, \citenamefont {Jordan}, \citenamefont
  {Graham},\ and\ \citenamefont {Kwiat}}]{dressel2013strengthening}%
  \BibitemOpen
  \bibfield  {author} {\bibinfo {author} {\bibfnamefont {J.}~\bibnamefont
  {Dressel}}, \bibinfo {author} {\bibfnamefont {K.}~\bibnamefont {Lyons}},
  \bibinfo {author} {\bibfnamefont {A.~N.}\ \bibnamefont {Jordan}}, \bibinfo
  {author} {\bibfnamefont {T.~M.}\ \bibnamefont {Graham}}, \ and\ \bibinfo
  {author} {\bibfnamefont {P.~G.}\ \bibnamefont {Kwiat}},\ }\href@noop {}
  {\bibfield  {journal} {\bibinfo  {journal} {Phys. Rev. A}\ }\textbf {\bibinfo
  {volume} {88}},\ \bibinfo {pages} {023821} (\bibinfo {year}
  {2013})}\BibitemShut {NoStop}%
\bibitem [{\citenamefont {Pryde}\ \emph {et~al.}(2005)\citenamefont {Pryde},
  \citenamefont {O'Brien}, \citenamefont {White}, \citenamefont {Ralph},\ and\
  \citenamefont {Wiseman}}]{Pryde2005}%
  \BibitemOpen
  \bibfield  {author} {\bibinfo {author} {\bibfnamefont {G.~J.}\ \bibnamefont
  {Pryde}}, \bibinfo {author} {\bibfnamefont {J.~L.}\ \bibnamefont {O'Brien}},
  \bibinfo {author} {\bibfnamefont {A.~G.}\ \bibnamefont {White}}, \bibinfo
  {author} {\bibfnamefont {T.~C.}\ \bibnamefont {Ralph}}, \ and\ \bibinfo
  {author} {\bibfnamefont {H.~M.}\ \bibnamefont {Wiseman}},\ }\href@noop {}
  {\bibfield  {journal} {\bibinfo  {journal} {Phys. Rev. Lett.}\ }\textbf
  {\bibinfo {volume} {94}},\ \bibinfo {pages} {220405} (\bibinfo {year}
  {2005})}\BibitemShut {NoStop}%
\bibitem [{\citenamefont {Hallaji}\ \emph {et~al.}(2017)\citenamefont
  {Hallaji}, \citenamefont {Feizpour}, \citenamefont {Dmochowski},
  \citenamefont {Sinclair},\ and\ \citenamefont {Steinberg}}]{Hallaji2017}%
  \BibitemOpen
  \bibfield  {author} {\bibinfo {author} {\bibfnamefont {M.}~\bibnamefont
  {Hallaji}}, \bibinfo {author} {\bibfnamefont {A.}~\bibnamefont {Feizpour}},
  \bibinfo {author} {\bibfnamefont {G.}~\bibnamefont {Dmochowski}}, \bibinfo
  {author} {\bibfnamefont {J.}~\bibnamefont {Sinclair}}, \ and\ \bibinfo
  {author} {\bibfnamefont {A.}~\bibnamefont {Steinberg}},\ }\href@noop {}
  {\bibfield  {journal} {\bibinfo  {journal} {Nat. Phys.}\ }\textbf {\bibinfo
  {volume} {13}},\ \bibinfo {pages} {540} (\bibinfo {year} {2017})}\BibitemShut
  {NoStop}%
\bibitem [{\citenamefont {Barnett}\ \emph {et~al.}(2003)\citenamefont
  {Barnett}, \citenamefont {Fabre},\ and\ \citenamefont
  {Ma{\i}tre}}]{barnett2003ultimate}%
  \BibitemOpen
  \bibfield  {author} {\bibinfo {author} {\bibfnamefont {S.~M.}\ \bibnamefont
  {Barnett}}, \bibinfo {author} {\bibfnamefont {C.}~\bibnamefont {Fabre}}, \
  and\ \bibinfo {author} {\bibfnamefont {A.}~\bibnamefont {Ma{\i}tre}},\
  }\href@noop {} {\bibfield  {journal} {\bibinfo  {journal} {The European
  Physical Journal D-Atomic, Molecular, Optical and Plasma Physics}\ }\textbf
  {\bibinfo {volume} {22}},\ \bibinfo {pages} {513} (\bibinfo {year}
  {2003})}\BibitemShut {NoStop}%
\bibitem [{\citenamefont {Byard}\ \emph {et~al.}(2014)\citenamefont {Byard},
  \citenamefont {Graham}, \citenamefont {Jordan},\ and\ \citenamefont
  {Kwiat}}]{byard2014increasing}%
  \BibitemOpen
  \bibfield  {author} {\bibinfo {author} {\bibfnamefont {C.}~\bibnamefont
  {Byard}}, \bibinfo {author} {\bibfnamefont {T.}~\bibnamefont {Graham}},
  \bibinfo {author} {\bibfnamefont {A.}~\bibnamefont {Jordan}}, \ and\ \bibinfo
  {author} {\bibfnamefont {P.}~\bibnamefont {Kwiat}},\ }in\ \href@noop {}
  {\emph {\bibinfo {booktitle} {2014 Conference on Lasers and Electro-Optics
  (CLEO)-Laser Science to Photonic Applications}}}\ (\bibinfo {organization}
  {IEEE},\ \bibinfo {year} {2014})\ pp.\ \bibinfo {pages} {1--2}\BibitemShut
  {NoStop}%
\bibitem [{\citenamefont {Kupchak}\ \emph {et~al.}(2019)\citenamefont
  {Kupchak}, \citenamefont {Erskine}, \citenamefont {England},\ and\
  \citenamefont {Sussman}}]{Kupchak2019}%
  \BibitemOpen
  \bibfield  {author} {\bibinfo {author} {\bibfnamefont {C.}~\bibnamefont
  {Kupchak}}, \bibinfo {author} {\bibfnamefont {J.}~\bibnamefont {Erskine}},
  \bibinfo {author} {\bibfnamefont {D.}~\bibnamefont {England}}, \ and\
  \bibinfo {author} {\bibfnamefont {B.}~\bibnamefont {Sussman}},\ }\href@noop
  {} {\bibfield  {journal} {\bibinfo  {journal} {Optics letters}\ }\textbf
  {\bibinfo {volume} {44}},\ \bibinfo {pages} {1427} (\bibinfo {year}
  {2019})}\BibitemShut {NoStop}%
\bibitem [{\citenamefont {Zhang}\ \emph {et~al.}(2018)\citenamefont {Zhang},
  \citenamefont {Zhang}, \citenamefont {Cen}, \citenamefont {Hu},\ and\
  \citenamefont {Zhao}}]{zhang2018simultaneously}%
  \BibitemOpen
  \bibfield  {author} {\bibinfo {author} {\bibfnamefont {J.-D.}\ \bibnamefont
  {Zhang}}, \bibinfo {author} {\bibfnamefont {Z.-J.}\ \bibnamefont {Zhang}},
  \bibinfo {author} {\bibfnamefont {L.-Z.}\ \bibnamefont {Cen}}, \bibinfo
  {author} {\bibfnamefont {J.-Y.}\ \bibnamefont {Hu}}, \ and\ \bibinfo {author}
  {\bibfnamefont {Y.}~\bibnamefont {Zhao}},\ }\href@noop {} {\bibfield
  {journal} {\bibinfo  {journal} {OSA Continuum}\ }\textbf {\bibinfo {volume}
  {1}},\ \bibinfo {pages} {1437} (\bibinfo {year} {2018})}\BibitemShut
  {NoStop}%
\bibitem [{\citenamefont {Fang}\ \emph {et~al.}(2020)\citenamefont {Fang},
  \citenamefont {Zhu}, \citenamefont {Jin}, \citenamefont {Tan}, \citenamefont
  {Li},\ and\ \citenamefont {Wu}}]{fang2020ultrasensitive}%
  \BibitemOpen
  \bibfield  {author} {\bibinfo {author} {\bibfnamefont {S.-Z.}\ \bibnamefont
  {Fang}}, \bibinfo {author} {\bibfnamefont {L.-L.}\ \bibnamefont {Zhu}},
  \bibinfo {author} {\bibfnamefont {R.-B.}\ \bibnamefont {Jin}}, \bibinfo
  {author} {\bibfnamefont {H.-T.}\ \bibnamefont {Tan}}, \bibinfo {author}
  {\bibfnamefont {G.-X.}\ \bibnamefont {Li}}, \ and\ \bibinfo {author}
  {\bibfnamefont {Q.-L.}\ \bibnamefont {Wu}},\ }\href@noop {} {\bibfield
  {journal} {\bibinfo  {journal} {Optics Communications}\ }\textbf {\bibinfo
  {volume} {460}},\ \bibinfo {pages} {125117} (\bibinfo {year}
  {2020})}\BibitemShut {NoStop}%
\bibitem [{\citenamefont {Lyons}\ \emph {et~al.}(2015)\citenamefont {Lyons},
  \citenamefont {Dressel}, \citenamefont {Jordan}, \citenamefont {Howell},\
  and\ \citenamefont {Kwiat}}]{lyons2015power}%
  \BibitemOpen
  \bibfield  {author} {\bibinfo {author} {\bibfnamefont {K.}~\bibnamefont
  {Lyons}}, \bibinfo {author} {\bibfnamefont {J.}~\bibnamefont {Dressel}},
  \bibinfo {author} {\bibfnamefont {A.~N.}\ \bibnamefont {Jordan}}, \bibinfo
  {author} {\bibfnamefont {J.~C.}\ \bibnamefont {Howell}}, \ and\ \bibinfo
  {author} {\bibfnamefont {P.~G.}\ \bibnamefont {Kwiat}},\ }\href@noop {}
  {\bibfield  {journal} {\bibinfo  {journal} {Phys. Rev. Lett.}\ }\textbf
  {\bibinfo {volume} {114}},\ \bibinfo {pages} {170801} (\bibinfo {year}
  {2015})}\BibitemShut {NoStop}%
\bibitem [{\citenamefont {Wang}\ \emph {et~al.}(2016)\citenamefont {Wang},
  \citenamefont {Tang}, \citenamefont {Hu}, \citenamefont {Wang}, \citenamefont
  {Yu}, \citenamefont {Zhou}, \citenamefont {Cheng}, \citenamefont {Xu},
  \citenamefont {Fang}, \citenamefont {Wu} \emph
  {et~al.}}]{wang2016experimental}%
  \BibitemOpen
  \bibfield  {author} {\bibinfo {author} {\bibfnamefont {Y.-T.}\ \bibnamefont
  {Wang}}, \bibinfo {author} {\bibfnamefont {J.-S.}\ \bibnamefont {Tang}},
  \bibinfo {author} {\bibfnamefont {G.}~\bibnamefont {Hu}}, \bibinfo {author}
  {\bibfnamefont {J.}~\bibnamefont {Wang}}, \bibinfo {author} {\bibfnamefont
  {S.}~\bibnamefont {Yu}}, \bibinfo {author} {\bibfnamefont {Z.-Q.}\
  \bibnamefont {Zhou}}, \bibinfo {author} {\bibfnamefont {Z.-D.}\ \bibnamefont
  {Cheng}}, \bibinfo {author} {\bibfnamefont {J.-S.}\ \bibnamefont {Xu}},
  \bibinfo {author} {\bibfnamefont {S.-Z.}\ \bibnamefont {Fang}}, \bibinfo
  {author} {\bibfnamefont {Q.-L.}\ \bibnamefont {Wu}},  \emph {et~al.},\
  }\href@noop {} {\bibfield  {journal} {\bibinfo  {journal} {Phys. Rev. Lett.}\
  }\textbf {\bibinfo {volume} {117}},\ \bibinfo {pages} {230801} (\bibinfo
  {year} {2016})}\BibitemShut {NoStop}%
\bibitem [{\citenamefont {Fang}\ \emph {et~al.}(2019)\citenamefont {Fang},
  \citenamefont {Zhu}, \citenamefont {Jin}, \citenamefont {Tan}, \citenamefont
  {Li},\ and\ \citenamefont {Wu}}]{fang2019ultrasensitive}%
  \BibitemOpen
  \bibfield  {author} {\bibinfo {author} {\bibfnamefont {S.-Z.}\ \bibnamefont
  {Fang}}, \bibinfo {author} {\bibfnamefont {L.-L.}\ \bibnamefont {Zhu}},
  \bibinfo {author} {\bibfnamefont {R.-B.}\ \bibnamefont {Jin}}, \bibinfo
  {author} {\bibfnamefont {H.-T.}\ \bibnamefont {Tan}}, \bibinfo {author}
  {\bibfnamefont {G.-X.}\ \bibnamefont {Li}}, \ and\ \bibinfo {author}
  {\bibfnamefont {Q.-L.}\ \bibnamefont {Wu}},\ }\href@noop {} {\bibfield
  {journal} {\bibinfo  {journal} {Opt. Commun.}\ }\textbf {\bibinfo {volume}
  {460}},\ \bibinfo {pages} {125117} (\bibinfo {year} {2019})}\BibitemShut
  {NoStop}%
\end{thebibliography}%

\end{document}